\documentclass[twocolumn,superscriptaddress,showpacs,amsmath,amssymb,pra,nofootinbib]{revtex4-1}

\usepackage{dcolumn,bm,book tabs,comment,natbib}
\usepackage[pdftex]{graphicx}
\usepackage[usenames,dvipsnames]{color}
\usepackage{colortbl}

\definecolor{URLCOL}{rgb}{0,0.52,0.83} 
\definecolor{LINKCOL}{rgb}{0.05,0.5,0} 
\definecolor{CITECOL}{rgb}{0.25,0,0.48} 

\usepackage[pdftex,bookmarks,breaklinks,bookmarksopen,bookmarksnumbered,colorlinks,linkcolor=LINKCOL,linktocpage,citecolor=CITECOL,urlcolor=URLCOL,pdfpagemode=UseOutline]{hyperref}

\definecolor{TITLECOL}{rgb}{0.1,0.2,0.7} 
\definecolor{SECOL}{rgb}{0.1,0.2,0.7} 
\definecolor{CONTENTSCOL}{rgb}{0.1,0.2,0.7} 
\definecolor{SSECOL}{rgb}{0.25,0,0.48} 
\definecolor{SSSECOL}{rgb}{0.2,0.08,0.53} 
\definecolor{FINCOL}{rgb}{0.01,0.3,0.07}

\def\coloredtitle#1{\title{\textcolor{TITLECOL}{#1}}} 
\def\coloredauthor#1{\author{\textcolor{CITECOL}{#1}}} 

\definecolor{URLCOL}{rgb}{0,0.17,0.43} 
\definecolor{LINKCOL}{rgb}{0.05,0.4,0} 
\definecolor{CITECOL}{rgb}{0.35,0,0.48} 

\definecolor{lightgray}{gray}{0.8}

\def\bea{\begin{eqnarray}}
\def\eea{\end{eqnarray}}
\def\ben{\begin{equation}}
\def\een{\end{equation}}
\def\benu{\begin{enumerate}}
\def\enu{\end{enumerate}}

\def\bei{\begin{itemize}}
\def\eei{\end{itemize}}
\def\beit{\begin{itemize}}
\def\eit{\end{itemize}}
\def\benu{\begin{enumerate}}
\def\enu{\end{enumerate}}

\def\n{n}

\def\sss{\scriptscriptstyle\rm}

\def\1var{(\bx_1...\bx\N)}

\def\half{\frac{1}{2}}

\def\br{{\bf r}}

\def\bx{{x}}

\def\x{_{\sss X}}
\def\c{_{\sss C}}
\def\ac{_{\sss AC}}
\def\s{_{\sss S}}
\def\xc{_{\sss XC}}

\def\N{_{\sss N}}

\def\LDA{^{\rm LDA}}
\def\RPA{^{\rm RPA}}
\def\QC{^{\rm QC}}
\def\GEA{^{\rm GEA}}
\def\GGA{^{\rm GGA}}
\def\nGGA{^{\rm RSC}}
\def\acGGA{^{\rm acGGA}}

\def\PBE{^{\rm PBE}}

\def\TF{^{\rm TF}}

\def\unif{^{\rm unif}}

\def\sph_int{ {\int d^3 r}}

\def\bx{\vectorsym{x}}

\newcommand{\LEXX}{\text{LEXX}}

\begin{document}
\coloredtitle{
Atomic correlation energies and the
generalized gradient approximation
}

\coloredauthor{Kieron Burke}
\affiliation{Department of Chemistry, 
University of California, Irvine, CA 92697}

\coloredauthor{Antonio Cancio}
\affiliation{Department of Physics and Astronomy, Ball State University,
Muncie, IN 47306}

\coloredauthor{Tim Gould}\affiliation{Qld Micro- and Nanotechnology Centre, %
Griffith University, Nathan, Qld 4111, Australia}

\coloredauthor{Stefano Pittalis}
\affiliation{CNR-Istituto di Nanoscienze, Via Campi 213A, I-41125 Modena, Italy}

\date{\today}

\begin{abstract}
Careful extrapolation of atomic correlation energies suggests that $E\c \to -A Z ln Z +  B Z$ as $Z \to \infty$, where $Z$ is the atomic number, $A$ is known, and $B$ is about 38 milliHartrees.  The coefficients roughly agree with those of the high-density limit of the real-space construction of the generalized gradient approximation.  An asymptotic coefficient, missed by previous derivations, is included in a revised approximation.  The exchange is also corrected, reducing atomic errors considerably.
\end{abstract}

\pacs{
71.15.Mb 
31.15.E- 
31.15.ve 
31.15.xp 
}

\maketitle


Modern density functional theory (DFT) is applied to a huge variety of
molecules and materials with many impressive results\cite{PGB14}.  But hundreds of
different approximations are available\cite{MOB12}, many of which contain empirical
parameters that have been optimized over some set of training data\cite{B93,LYP88}.
The most popular non-empirical approximations are those of Perdew and co-workers,
which eschew empiricism in favor of exact conditions using only the
uniform and slowly-varying electron gases for input. 
But even this
non-empirical approach can 
require judicious choice among exact conditions\cite{HTBP10},
which can appear
at odds with claims of DFT being
first-principles\cite{PGB14}.

A unified, systematic approach to functional approximation {\em is} possible.
Lieb and Simon\cite{LS73} showed that the ground-state energy in  
Thomas-Fermi (TF) theory
becomes relatively exact in a specific high-density, large particle number
limit.  In a peculiar sense
the density, $\n(\br)$,
also approaches that of TF 
\cite{L81}.  
For model systems, the leading corrections to TF, derived semiclassically,
have been shown to be much more 
accurate than typical density functional approximations\cite{ELCB08}.
The simplest example of this limit is $Z\to\infty$ for neutral atoms,
where
the local density approximation (LDA) to the exchange energy, $E\x$,
becomes relatively exact\cite{S81}.   Some
modern generalized gradient approximations (GGA's)
yield the leading energetic correction\cite{PCSB06,EB09}.
While such limits themselves do not wholly determine approximations,
they do indicate precisely {\em which} limits a non-empirical approximation must
satisfy\cite{HTBP10}.  Understanding of this limit was key to the PBEsol
approximation that was designed to improve lattice parameters in solids\cite{PRCV08}.
\begin{figure}[t]
\includegraphics[width=0.9\columnwidth]{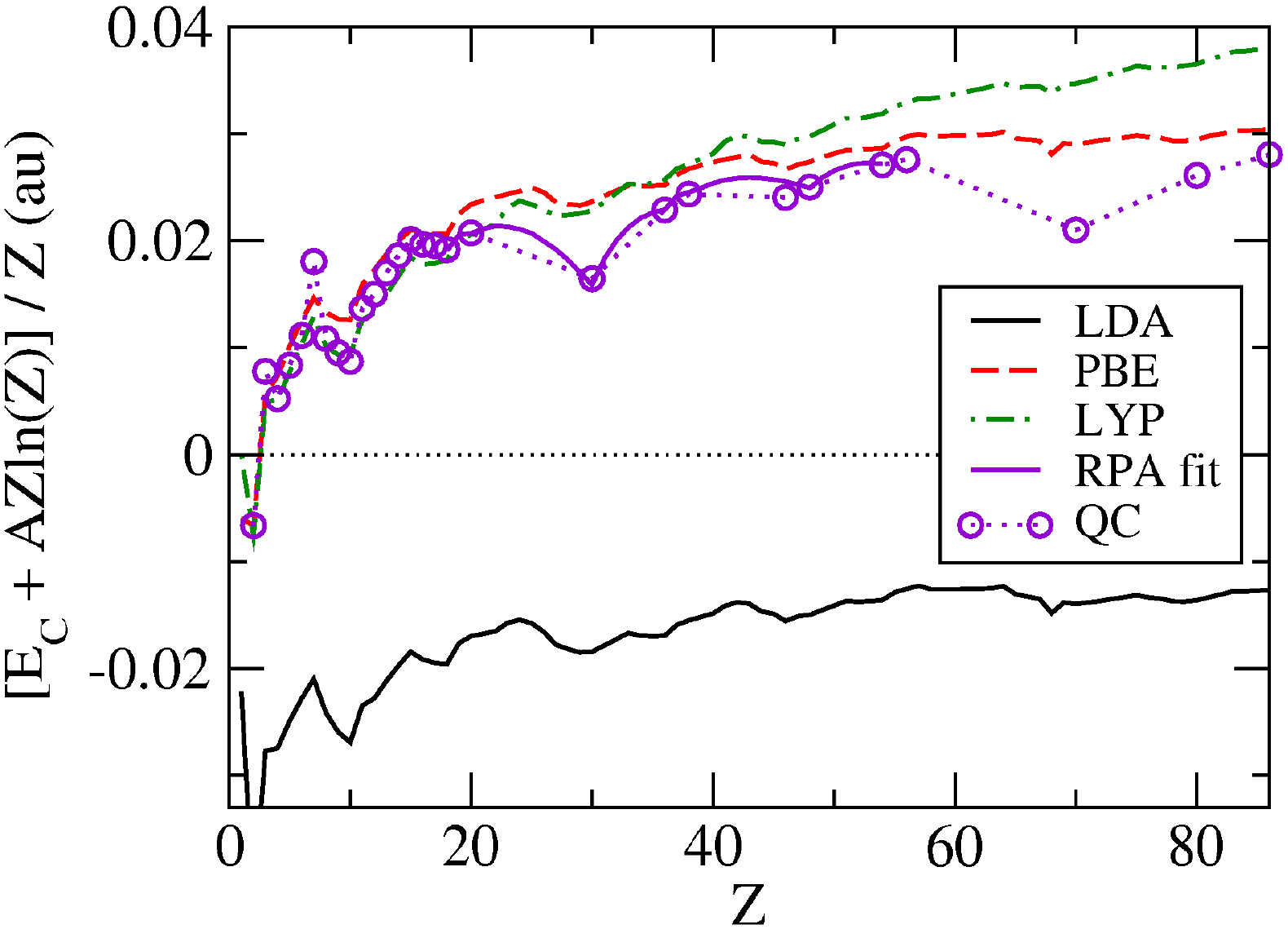}
\caption{
Atomic correlation energies per electron, with the leading large-$Z$
contribution removed.
Accurate quantum chemical (QC) results (see text)
and approximate self-consistent DFT results.}
\label{apEcall}
\vskip -0.5cm
\end{figure}

Here, we extend this idea to correlation, and ask:  Does LDA yield the
dominant term, and should GGA recover the leading correction?
From GGA, we derive a simple formula for the large-$Z$ correlation
energy of atoms.
Reference quantum chemical (QC) energies for spherical atoms\cite{MT11}
match this form, and asymptotic coefficients can be extracted numerically.
We find that the real-space cutoff construction of a GGA roughly
reproduces this number, validating both that procedure\cite{BPW97c} and
the seminal idea of Ma and Bruckner\cite{MB68}.
Fig.~\ref{apEcall} shows
non-empirical GGAs like PBE are highly accurate
in this limit, while empirical formulas such as LYP\cite{LYP88} fail.

Taking advantage of this insight, we explore the behavior of the
GGA at large $Z$, finding an important new condition for correlation.
A new approximation is created that satisfies this condition.
To test this form, we fill in atomic correlation energies over
more of the periodic table, by performing
random phase appoximation (RPA) 
calculations for non-spherical atoms up to Xe, and correct
them to yield accurate correlation energies.
The errors of the new approximation across periods are consistent with its
semiclassical derivation, and are smaller by a factor of 2 than those of PBE.
A related correction to the PBE exchange then yields errors smaller by a factor of 4.
These results 
indisputably tie the uniform and slowly-varying
gases to real systems.

We begin our analysis with the uniform electron gas (jellium).
In a landmark of electronic structure theory, Gell-Mann and Brueckner\cite{GB57}
applied the
random phase approximation (RPA) to find:
\ben
\epsilon\c\unif = c_0\ \ln r_s - c_1 + ...,~~~r\s\to 0
\label{epsunif}
\een
where $r\s=(3/(4\pi\n))^{1/3}$ is the Wigner-Seitz radius of density $\n$, $c_0=0.031091$,
and $c_1\RPA=0.07082$.  We use atomic units (energies in 
Hartrees) and give derivations
for spin-unpolarized systems
for simplicity, but all calculations include spin-polarization, unless otherwise
noted.  Our aim is to find
the non-relativistic limit and all results are for this case.
In fact, Eq. (\ref{epsunif}) yields the exact high-density limit if
$c_1=0.04664$, 
the correction from RPA being due to second-order exchange.
An accurate modern parametrization that contains
these limits is given in Ref. \cite{PW92}.
Then\cite{KS65}
\ben
E\c\LDA[\n] = \int d^3r\, \n(\br)\ \epsilon\c\unif(\n(\br)),
\label{EcLDA}
\een
which greatly overestimates the magnitude of the correlation energy of atoms
(factor of 2 or more).
For atoms with large $Z$, insert $\n\TF(\br)$ into Eq. (\ref{EcLDA}) to find:
\ben
E\c\LDA = - A Z\, \ln Z\, + B\LDA Z + ...,
\een
where $A=2 c_0/3=0.02072$ and $B\LDA =-0.00451$.  The first term is
exact for atoms\cite{KR10}, so we define 
\ben
B = \lim_{Z\to\infty} (E\c(Z)/Z+A \ln Z).
\label{Bdef}
\een
Fig. 1 suggests $B$ is finite, undefined in LYP, finite
but inaccurate in LDA, and roughly correct in PBE.

To understand why PBE should be accurate, we review the history of non-empirical
GGAs.
Again within RPA, Ma and Brueckner (MB) 
derive the leading gradient correction for the correlation energy of
a slowly varying electron gas\cite{MB68}.  Defining 
\ben
\Delta E\c = E\c - E\c\LDA,
\een
the gradient expansion
approximation yields
\ben
\Delta \epsilon\c\GEA(\br) = \beta\, t^2(\br),~~~~~(r\s\to 0)
\een
where $t= |\nabla \n|/(2 k_s \n)$ is the dimensionless gradient
for correlation, $\beta=0.066725$ and $k\s=2 (3\n/\pi)^{1/6}$ is the TF screening
length\cite{PBE96}. This
so strongly overcorrects $E\c\LDA$ for atoms that some $E\c$ become positive.
MB showed that a simple Pade approximant
works much better, creating the 
first modern GGA, and
inspiring the work of Langreth and Perdew\cite{LP80}, among others.

But underlying some GGAs is the 
non-empirical real-space cutoff (RSC) procedure  for the XC
hole\cite{BPW97c}.  Write
\ben
E\xc = \half\int d^3r\, \int d^3r'\, \frac{\n(\br)\, \n\xc(\br,\br')}{|\br-\br'|}.
\label{Exchole}
\een
The LDA can be considered as approximating the true XC hole by that of a
uniform gas:
\ben
\n\xc\LDA (\br,\br') = \n(\br)\, (\bar g\unif(r\s(\br),|\br-\br'|) -1)
\label{nxcLDA}
\een
where $\bar g\unif$ is the (coupling-constant averaged) pair-correlation function
of the uniform gas\cite{PW92b}.  
Insertion of this approximate hole into Eq. (\ref{Exchole})
yields $E\xc\LDA[\n]$.
While $e\xc\unif(\n(\br))$ is not accurate
point-wise\cite{JG89}, the system- and spherical average of the LDA hole is.
This is because the LDA hole satisfies some basic conditions (normalized
to -1 and $n\x(\br,\br') \leq 0$), so it roughly mimics
the exact hole.   Hence the reliability and systematic errors of LDA
\cite{JG89}.

This analysis shows why the gradient expansion fails:
$\n\xc\GEA$ for a 
non-slowly varying system has large unphysical
corrections to $\n\xc\LDA$, 
violating the exact conditions\cite{BPE98}.  The RSC
construction sharply cuts off the
parts of the hole that violate these conditions.  
The PBE functional is a parametrization of RSC and
the paper also showed how the basic features could be deduced by restraining
simple forms with exact conditions\cite{PBE96}.  Its form for correlation is
\ben
\Delta \epsilon\c\PBE(r\s,t)= c_0\, \ln \left[
1 +\beta\, t^2\, F(\tilde A t^2)/c_0\right],
\label{Hpbe}
\een
where $F=(1+x)/(1+x+x^2)$, and
\ben
\beta \tilde A^{-1} = c_0\left[ \exp (-\epsilon\c\unif/c_0) -1\right].
\label{Adef}
\een
This form yields a finite $E\c$ in the high-density limit of
finite systems, zero correlation as $t\to\infty$, and 
recovers the gradient expansion for small $t$, just as RSC does\cite{PBE96}.

\begin{figure}[t]
\includegraphics[width=0.9\columnwidth]{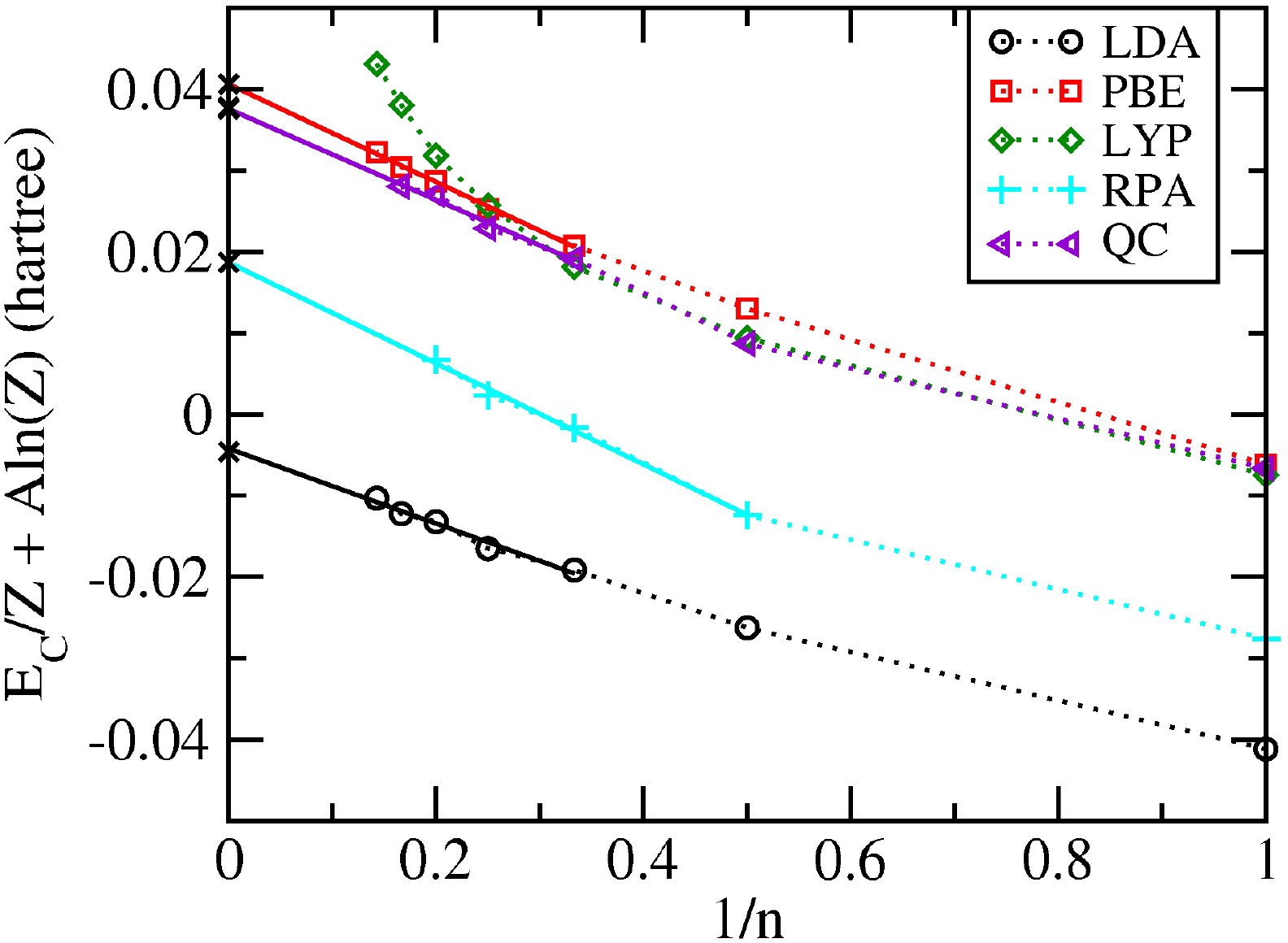}
\caption{Same as Fig 1, but
plotted against inverse principal quantum number, and using only noble gases.
Solid lines are straight-line extrapolations to estimate $B$
of Eq. (\ref{Bdef}).
}
\label{Ecasy}
\vskip -0.5 cm
\end{figure}
Everything discussed so far is long known.  Our analysis begins by noting
that accurate values of $B$ cannot be extracted directly from Fig. 1.
The extrapolation to $Z\to\infty$ is obscured by the oscillations
across the periodic table.
Using methods developed in Ref. \cite{CSPB10},
in Fig. 2 we include only noble gases and
plot as a function of inverse principal
quantum number.  By linear extrapolation, we find $B$ is about
0.038 for QC, and 0.0406 for PBE.
The analogous plots for alkali earths give asymptotes 
1 to 2 milliHartree higher.
We can also find $B\PBE$ analytically.  As $r\s\to 0$,
$\tilde{A}\to 0$ and $F\to 1$, yielding
\ben
\Delta \epsilon\c\PBE(0,t) = c_0\, \ln
( 1 +\beta t^2/c_0). 
\label{H0pbe} 
\een
Inserting an accurate\cite{LCPB09} $\n\TF(\br)$ yields
$B\PBE=0.0393$ exactly\cite{PCSB06}.  
The agreement within
a few milliHartree validates the extrapolation.

We pause to discuss the message of Figs 1 and 2.  
First, the original idea of MB, that of resumming the gradient
expansion, is validated, but 
the real-space construction for the correlation hole is needed.
Second, the LDA and RSC determine the correlation energies of atoms for
large $Z$, explaining their relevance to atomic and molecular systems.
Third, the large-$Z$ expansion determines
which conditions in functional construction ensure accurate energies.
Any approximation, such as LYP, which does not produce the $AZ\ln Z$ dependence,
worsens with increasing $Z$:   LYP is not optimal
even for 3d transition metal complexes\cite{FP06}.

\begin{figure}[htb]
\includegraphics[width=0.9\columnwidth]{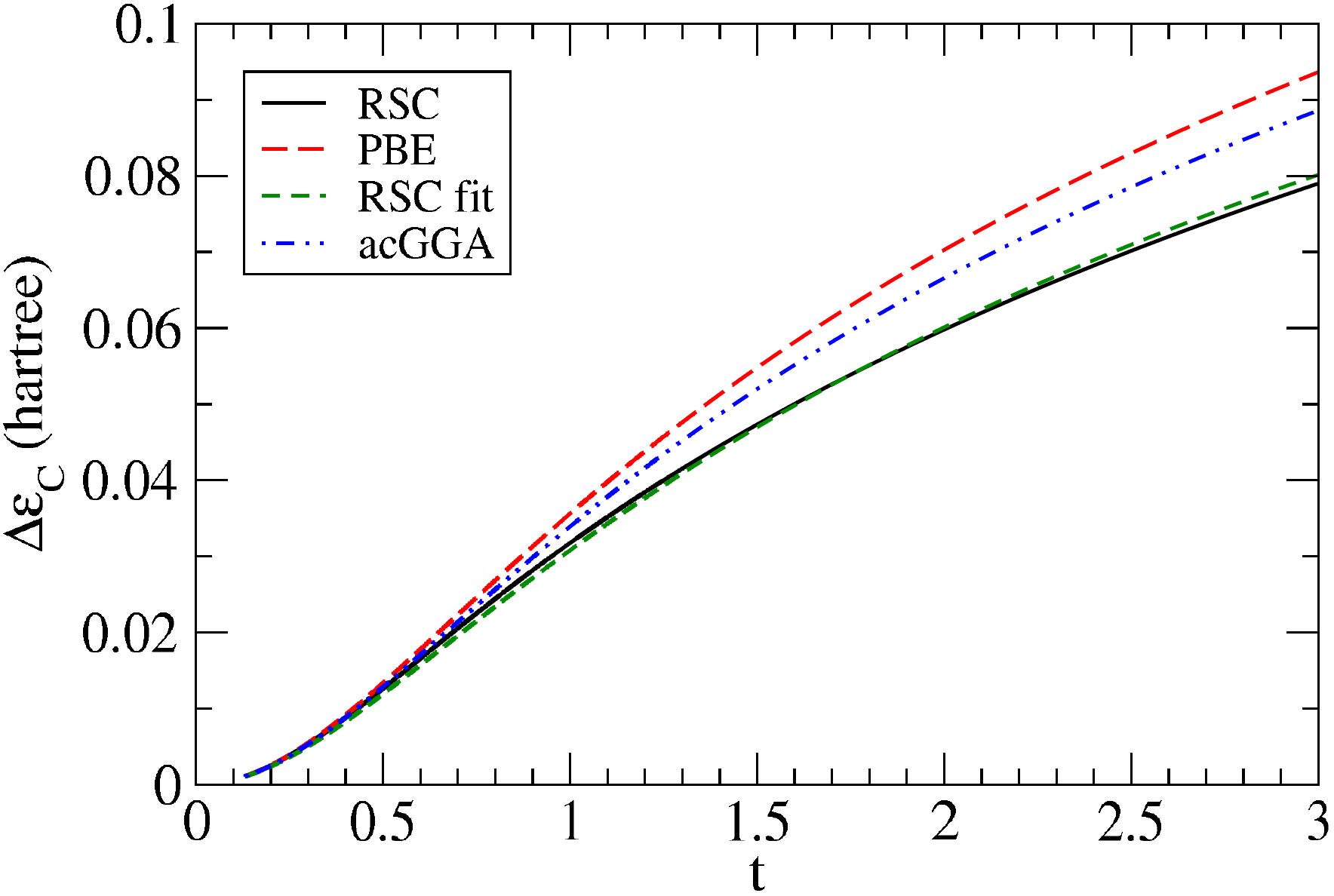}
\caption{
$\Delta \epsilon\c(0,t)$ for real-space cutoff (RSC), PBE, our fit to RSC,
and the asymptotically correct GGA (acGGA).
}
\label{HGGA}
\vskip -0.3 cm
\end{figure}
The accuracy of $B\PBE$
suggests the real-space
cut-off procedure is highly accurate here.
To check this, we derive RSC in the large-$Z$ limit.
Appendix C of Ref. \cite{BPW97c} gives formulas for RSC as $r\s\to 0$.  
Solving the RSC
equations numerically, we find Fig.~\ref{HGGA}, which
compares PBE and RSC for the high-density
limit.  
The result is rather surprising.  Although PBE follows RSC almost perfectly
for $t < 1/2$, they clearly differ for large $t$.  They both have the same
$\ln t$ divergence for large $t$, but differ in the next order.  Define
\ben
C=\lim_{t\to\infty}(\Delta \epsilon\GGA\c(0,t) - 2 c_0\, \ln t),
\een
to find $C\PBE= c_0 \ln(\beta/c_0) =0.0237$.   For the real-space construction,
define
\ben
\gamma = \lim_{\epsilon\to 0} \int_\epsilon^\infty dv\, \frac{f_1(v)-4c_0}{2v},
\een
where $v=k\s u$ and $f_1(v)$ is the dimensionless 
radial $\n\c\LDA(u)$ in RPA.  Then
\ben
C\nGGA = c_0 ( 3 - 2 \ln (3\pi{\sqrt{6 c_0}})) + \gamma,
\een
which is about -0.0044 with
the models of Ref. \cite{BPW97c}.  We use this
to fit the RSC curve with a simple form:
\ben
\Delta \epsilon\c\nGGA(0,t) = c_0\, \ln \left[ 1 + \beta t^2 P(t) / c_0 \right]
\een
where $P(t) = (1 + t/\tau)/(1+ \tilde c t/\tau)$, $c_0\ln\tilde c= C\PBE-C\nGGA$,
and $\tau=4.5$ is chosen to match the RSC curve.  This
satisfies all the conditions of PBE correlation, plus one more: it contains
the RSC leading
correction to $\ln t$, unlike PBE.   Then $B\nGGA = 0.0327$.
The 
difference from PBE is small,
because $t$ 
is $< 1$ for most of the TF atom (see Fig 9 of Ref. \cite{LCPB09}),
reflecting
the uncertainty in RSC in this limit.

To construct an approximation without this uncertainty, we
keep $\tau$ the same, but choose $\tilde c\ac \!=\! 1.467$, which reproduces
our best estimate of $B$.
Our asymptotically 
corrected GGA (acGGA) is then
\ben
\Delta \epsilon\c^{\rm acGGA}(r_s,t) = 
    c_0 \, \ln \left[ 1 + \beta \tilde{t}^2 F(\tilde{t}^2) / c_0 \right].
\een
where $\tilde{t} = t \sqrt{P(t)}$.
acGGA satisfies {\em all} the conditions of PBE correlation, but also reproduces
the correct large-$Z$ behavior for atoms.
Unlike other suggestions\cite{ZY98,PRCV08},
this is not a change of the constants in the PBE form, but a new form without
which this condition cannot be satisfied.

We need more data to test and understand our
approximation: $Z$ up to 18\cite{CGDP93} is too far from the asymptotic region, and the
spherical atoms alone\cite{MT11} are too sparse.
For $19 \leq Z\leq 54$ we have performed RPA calculations, evaluating
the coupling constant integration
using the fluctuation-dissipation theorem
at imaginary frequency.
We used the optimized effective potential $v_{\LEXX}(\br)$
for the linear exact exchange (LEXX)\cite{Gould2013-LEXX} functional
(found via the Krieger, Li and Iafrate approximation\cite{KLI1992}).
Even for open-shell systems $v_{\LEXX}(\br)$ is independent
of both spin and angle, 
and includes important features of the
exact potential due to its inclusion of static correlation.
Details can be found in
Refs.~\onlinecite{Gould2013-LEXX,Gould2013-Aff}, and
its extension to $d$ shells via ensemble averaging\cite{Perdew1982} 
will be discussed in a longer paper.
We then fit the correction:
\ben
E\QC \approx E\RPA + Z(0.0199 +0.00246/n(Z) + ...)
\een
where $n(Z)=(6Z+8)^{1/3} - 2$.
This agrees almost exactly with
spherical atoms in the range $19 \leq Z \leq 54$
{(Fig.~\ref{apEcall}), and with all atoms in $11\leq Z \leq 18$,
another illustration of the power of asymptotic analysis.}
All energies are listed in Supplementary Information.

\begin{table}[h]
\begin{tabular}{|c|c|c|c|c|c||c|c|c|}
\hline
 & \multicolumn{5}{c||}{$E\c$} & \multicolumn{3}{c|}{$E\xc$}\\
\hline
p & RPA+ & LDA & LYP & PBE & ac & PBE & b88-ac &ac \\
\hline \hline
1 & 0.035 & 0.765 & 0.012 & 0.081 & 0.094 & 0.216 & 0.038 & 0.034\\
2 & 0.074 & 0.941 & 0.038 & 0.069 & 0.053 & 0.287 & 0.071 & 0.032\\
3 & 0.020 & 1.032 & 0.045 & 0.041 & 0.013 & 0.297 & 0.024 & 0.103\\
4 & 0.010 & 1.000 & 0.085 & 0.111 & 0.062 & 0.357 & 0.018 & 0.113\\
5 & 0.020 & 1.087 & 0.103 & 0.049 & 0.008 & 0.428 & 0.009 & 0.087\\
\hline \hline
all & 0.026 & 1.019 & 0.089 & 0.074 & 0.038 & 0.365 & 0.025 & 0.086\\
\hline 
\end{tabular}
\caption{Mean absolute error (eV) of energy components per electron,
taken with respect to our
reference data set,
and averaged over each period (p) of the periodic table.}
\label{XCmae}
\vskip -0.25cm
\end{table}
The left side of Table \ref{XCmae} lists average errors for atomic correlation
with respect to this reference set.  
LDA overestimates by about 1 eV per electron, consistent with its error for $B$.
PBE reduces this error by about a factor of 10, consistent with its (almost) exact
value for $B$. acGGA reduces this error by a further factor of 2, by being exact for $B$.
The nefarious LYP does best for $Z < 10$, vital to organic chemistry, but is far worse
past period 3.  
Even RPA+\cite{KP99}, which requires RPA correlation energies, is only slightly better
than acGGA.

\begin{figure}[htb]
\includegraphics[width=0.85\columnwidth]{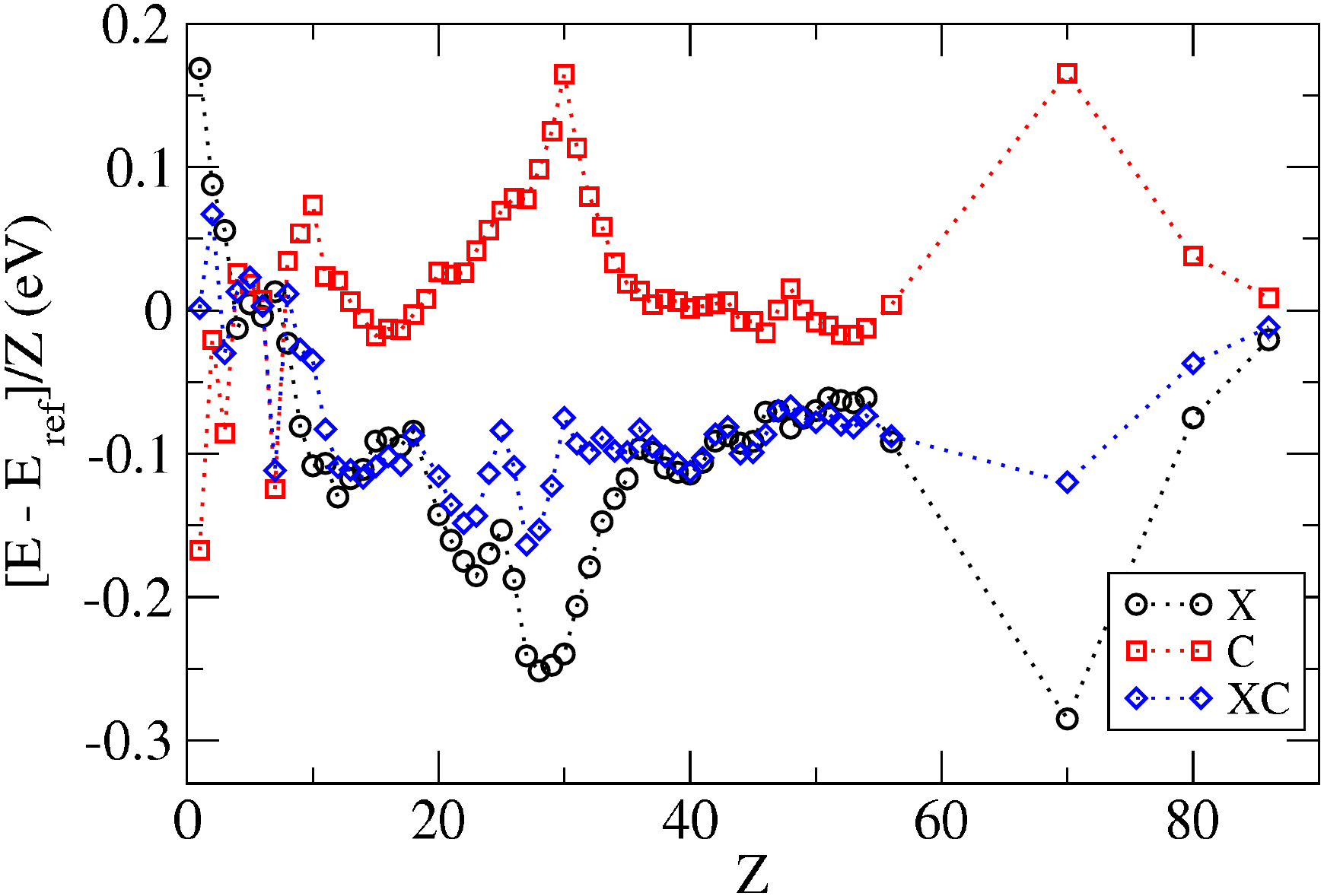}
\caption{Errors in energy components per electron as a function
of $Z$ for acGGA for X, C, and XC together.  PBE errors are
significantly larger (see Table \ref{XCmae}) and do not often cancel.}
\vskip -0.3cm
\label{Eax}
\end{figure}
We close by applying the same methodology to $E\x$.  In Ref. \cite{EB09}, it was shown that
both B88\cite{B88} and PBE are asymptotically accurate for $E\x$.
But close inspection of Table 1 of Ref. \cite{EB09}
shows a small underestimate in the coefficient from PBE.  To correct this, we 
simply increase $\mu$ in the formula for $E\x\PBE$ by 13\%, to 0.249.
We combine this with our correlation approximation to make acGGA, and plot XC energy errors
per electron in Fig. \ref{Eax}.  
Remarkably, the bad actors ($Z=10,30,70$)
are {\em the same} for both X and C.  
To understand why these are bad, note
that the semiclassical expansion performs worst when only the lowest level of a spatial
orbital is occupied\cite{CLEB10}.   
Writing $\n(\br)$ as a sum over contributions from different $l$-values,
the bad actors are those at which a new $l$-shell
has been filled for the first time (2p, 3d, 4f, respectively).  
A small energy error per electron
keeps adding until that shell is filled.  
No such error occurs in the following period, so that even periods
have much large acGGA errors than odd periods in Table 1.
The X and C errors are
like mirror images, so that they somewhat cancel each other, just as in LDA.
This is reflected in the right side of Table~\ref{XCmae}.
For XC together, the acGGA MAE is less than a quarter of PBE's.
In fairness, we note that the empirical B88 exchange functional is so asymptotically accurate
that, combined with acGGA correlation, its error is three times smaller again.

Our results are quite general.  Lieb-Simon scaling can be applied
to any system, if bond lengths are also scaled\cite{L81}.  TF energies
become relatively exact, and we believe LDA X and C both do too.
We expect $B\acGGA$ to be highly accurate for molecules and solids,
so our acGGA for XC may well prove more accurate than PBE for more than just
atoms.

KB and SP thank NSF CHE-1112442 for support.  
SP acknowledges funding by the European Commission (Grant No. FP7-NMP-CRONOS). 
TG recognises computing support from the Griffith University Gowonda
HPC Cluster.
No empirical parameters were used in the construction of the approximation
suggested here.  Explicit formulas are given in Supplemental Information,
and a more detailed account is in preparation.

%



\end{document}